\documentclass{PoS}

\usepackage{amsmath}

\def\apj{Astrophys.~J.}%
\def\apjl{Astrophys.~J.~Letters}%
\def\jcap{J.~Cosmol.~Astropart.~Phys.}%
\def\mnras{Mon. Not. R. Astron. Soc.}%
\def\prd{Phys.~Rev.~D}%
\def\prl{Phys.~Rev.~Lett.}%

\newcommand{\mrm}[1]{\mathrm{#1}}

\title{Self-interacting Dark Matter Energy Density}

\ShortTitle{Self-interacting Dark Matter Energy Density}

\author{Rainer Stiele, Tillmann Boeckel, J\"urgen Schaffner-Bielich\\
        Institute for Theoretical Physics, Heidelberg University,\\ Philosophenweg 16, D-69120 Heidelberg, Germany\\
        E-mail: \email{r.stiele@thphys.uni-heidelberg.de}}

\abstract{We investigate cosmological implications of an energy density contribution arising by elastic dark matter self-interactions. Its scaling behaviour shows that it can be the dominant energy contribution in the early universe. Constraints from primordial nucleosynthesis give an upper limit on the self-interaction strength which allows for the same strength as standard model strong interactions. Furthermore we explore the cosmological consequences of an early self-interaction dominated universe. Chemical dark matter decoupling requires that self-interacting dark matter particles are rather light (keV range) but we find that super-weak inelastic interactions are predicted by strong elastic dark matter self-interactions. Assuming a second, collisionless cold dark matter component, its natural decoupling scale exceeds the weak scale and is in accord with the electron and positron excess observed by PAMELA and Fermi-LAT. Structure formation analysis reveals a linear growing solution during self-interaction domination, enhancing structures up to $\sim 10^{-3} M_\odot$ long before the formation of the first stars.}

\FullConference{25th Texas Symposium on Relativistic Astrophysics \\
                 December 6-10, 2010\\
                 Heidelberg, Germany}

\begin{document}

\section{Self-interacting Dark Matter}
	Finite elastic interactions between dark matter particles can be constrained by their various implications on the dark matter distribution. Concerning dark matter halos, substructures can get evaporated \cite{Gnedin_2001} and spherical cores are formed \cite{Dave_2001, Yoshida_2000, Miralda_2002} due to the energy exchange. Bounds on the drag force between dark matter structures are found from cluster collisions \cite{Randall_2008}. But there are also very recent indications for finite self-scattering of dark matter by the spatial separation between the stellar and dark matter components of galaxies in clusters, analysed in Ref.~\cite{Williams_2011}. All these limits are summarised in Table~\ref{SIcs_limits}. The last column gives the corresponding bounds on the interaction energy scale used in Sec.~\ref{sec:rho_SI} according to Eq.~(\ref{eq:sigmaSI}). Interaction strengths of the order $m_\mrm{SI}\,/\!\sqrt{\alpha_\mrm{SI}}\sim10 - 100\,\mrm{MeV}$ are consistent with all observations.
	
	So it is worthwhile to investigate further implications well in these limits. Here, we present an overview based upon Ref.~\cite{Stiele_2010} of the evidence and the implications of an interaction energy density contribution. Other recent investigations of finite elastic dark matter self-interactions are given in Refs.~\cite{Feng_2009, Buckley_2010}.
	 
	 \begin{table}[h]
		\begin{center}
			\begin{tabular}{c|c|c|c}
				& $\sigma_\mrm{SI}/m_\mrm{SIDM}\,\left[\mrm{cm}^2/\mrm{g}\right]$ & Ref. & \begin{minipage}[l]{3cm} $m_\mrm{SI}\,/\!\sqrt{\alpha_\mrm{SI}}\hskip1.5ex\left[\mrm{MeV}\right] $ \\ $/\left(m_\mrm{SIDM}/1\,\mrm{keV}\right)^{1/4}$ \end{minipage} \\
				\hline
				Galactic evaporation &  $\lesssim 0.3 $ & \cite{Gnedin_2001} & $\gtrsim 7.4$\\ 
				Core sizes & $\lesssim 0.56 - 5.6$  & \cite{Dave_2001, Yoshida_2000} & $\gtrsim 3.5 - 6.3$\\
				Cluster ellipticity & $\lesssim 0.02 $ & \cite{Miralda_2002} & $\gtrsim 14.5$\\ 
				Bullet cluster & $< 0.7 - 1.25 $ & \cite{Randall_2008} & $> 5.1 - 5.9$\\
				Light/Mass separation & $\gtrsim 4.5\times10^{-7} - 0.05 $ & \cite{Williams_2011} & $\lesssim 11.5 - 210$\\
			\end{tabular}
			\caption{Bounds on the dark matter self-scattering cross section from halo properties if self-interacting dark matter is the only dark matter component. The last column gives the corresponding limits on the interaction energy scale used in Sec.~2 according to Eq.~(2.3).}
		\end{center}
		\label{SIcs_limits}
	\end{table}

\section{\label{sec:rho_SI}Self-interaction Energy Density}
\begin{figure}
		\centering
 		\includegraphics[width=0.75\textwidth]{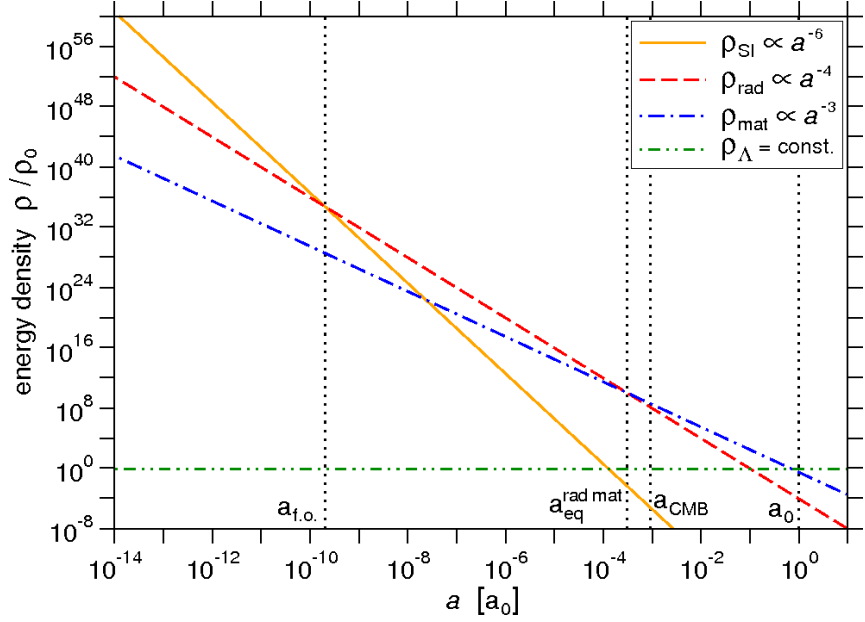}
		\caption[]{The evolution of different energy density contributions with the scale factor $a$. $a_\mrm{CMB}$ is the scale factor at photon decoupling, $a_\mrm{eq}^\mrm{rad\,mat}$ at radiation-matter equality and $a_\mrm{f.o.}$ at the freeze-out of the neutron to proton number ratio. $\varrho_\mrm{SI}$ is fixed so that self-interaction--radiation equality is at $a_\mrm{f.o.}$. The universe could be in a self-interaction dominated epoch prior to radiation domination in the very early universe.}
		\label{fig:rhos_scaling}
	\end{figure}

	To quantify the elastic dark matter self-interaction (SI) we consider the simplest model of two-body interactions. In a lowest order approximation the interaction energy density is proportional to $n_\mrm{SIDM}^2$, where $n_\mrm{SIDM}$ is the self-interacting dark matter (SIDM) number density. To have the correct dimensionality this term can be written as $\varrho_\mrm{SI}=n_\mrm{SIDM}^2\, / \!\left(m_\mrm{SI}^2/\!\alpha_\mrm{SI}\right)$, where $m_\mrm{SI}\,/\!\sqrt{\alpha_\mrm{SI}}$ represents the energy scale of the interaction. 
	The corresponding contribution to the pressure is
	\begin{equation}
		\varrho_\mrm{SI} = \frac{\alpha_\mrm{SI}}{m_\mrm{SI}^2}\,n_\mrm{SIDM}^2 = p_\mrm{SI}\;.
		\label{eq:eos_SI}
	\end{equation}
	For comparison two standard model interactions are used. For weak interactions the interaction strength is $m_\mrm{weak}\,/\!\sqrt{\alpha_\mrm{weak}}\sim300\,\mrm{GeV}$ and for strong interactions $m_\mrm{strong}\,/\!\sqrt{\alpha_\mrm{strong}}\sim100\,\mrm{MeV}$. The interaction is repulsive by construction which avoids an enhancement of the annihilation cross-section due to the formation of bound states.
	For the field theoretical derivation we refer to Ref.~\cite{Stiele_2010}. References \cite{Narain_2006, Agnihotri_2009} explored implications of this self-interaction energy density contribution on the mass-radius relation of compact stars made of self-interacting dark matter.\pagebreak\\
	The equation of state (\ref{eq:eos_SI}) as input to the Friedman equations determines the scaling behaviour of the self-interaction energy density with the scale factor $a$:
	\begin{equation}
		\varrho_\mrm{SI}\propto{a^{-6}}\;.
		\label{eq:rhoSI_a}
	\end{equation}
	Hence, $\varrho_\mrm{SI}$ shows the steepest possible decrease and thus a steeper  decrease than the standard ingredients of the universe. The universe could be in a self-interaction dominated epoch prior to radiation domination in the very early universe. The evolution of the different energy density contributions is illustrated in Fig.~\ref{fig:rhos_scaling}.\\
	The scaling relations (\ref{eq:eos_SI}) and (\ref{eq:rhoSI_a}) imply that $n_\mrm{SIDM}\propto{a^{-3}}$ during self-interaction domination. This is true for decoupled or relativistic particles, in any case relatively light particles. The recently highlighted dark matter particle mass at the keV scale \cite{deVega_2010} is an attractive candidate for self-interacting dark matter.\\
		A relationship between the interaction energy scale and the self-scattering cross section can be given by
	\begin{equation}
		\sigma_\mrm{SI}\approx{s\,\frac{\alpha_\mrm{SI}^2}{m_\mrm{SI}^4}} \quad \Leftrightarrow \quad \frac{m_\mrm{SI}}{\sqrt{\alpha_\mrm{SI}}}\approx5.44\,\mrm{MeV}\times\left(\frac{E_\mrm{SIDM}^2/m_\mrm{SIDM}}{1\,\mrm{keV}}\frac{1\,\mrm{cm^2/g}}{\sigma_\mrm{SI}/m_\mrm{SIDM}}\right)^{1/4}\;,
		\label{eq:sigmaSI}
	\end{equation}
	where $s=4E_\mrm{SIDM}^2$ in the center of momentum frame, with $E_\mrm{SIDM}\sim{T_\mrm{SIDM}}\,\left(\ \sim{m_\mrm{SIDM}}\right)$ as the \mbox{(non-)}relativistic single-particle energy.
	
\section{Constraints from Primordial Nucleosynthesis}
	Primordial nucleosynthesis is the physical process of choice to constrain the self-interaction strength. Any additional energy density contribution increases the expansion rate $H\propto{\varrho^{1/2}}$ on which the resulting element abundances depend. According to Eq.\ (\ref{eq:rhoSI_a}) the relative contribution of the self-interaction energy density is largest at the earliest stage of primordial nucleosynthesis, the freeze-out of the neutron to proton number ratio. Nearly all neutrons available for the nucleosynthesis processes are incorporated into $^4\mrm{He}$, so the final $^4\mrm{He}$ abundance is most sensitive to the expansion rate when the neutron/proton ratio freezes out. We can translate the primordial $^4\mrm{He}$ abundance $Y_\mrm{P}$ from observations into the following constraint on the dark matter self-interaction strength:
	\begin{equation}
		Y_\mrm{P}\simeq0.256\ \cite{Iz_Av_2010} \quad \Leftrightarrow \quad \frac{m_\mrm{SI}}{\sqrt{\alpha_\mrm{SI}}}\gtrsim1.70\,\mrm{keV}\times\frac{F_\mrm{SIDM}^0}{m_\mrm{SIDM}/1\,\mrm{keV}}\;.
	\label{limit_BBN}
	\end{equation}
	The relative amount of self-interacting dark matter $F_\mrm{SIDM}^0\equiv\Omega_\mrm{SIDM}^0/\Omega_\mrm{DM}^0$ serves to include the possibility of multiple dark matter components. Even an additional energy density contribution of dark matter self-interactions of the strength of the strong interaction ($m_\mrm{strong}/\!\sqrt{\alpha_\mrm{strong}}\sim100\,\mrm{MeV}$) is consistent with the primordial element abundances.\\
	Compared to the constraints given in Table~\ref{SIcs_limits} the constraint derived from primordial nucleosynthesis (\ref{limit_BBN}) is compatible only for the smallest dark matter particle masses. Nevertheless it gives a limit from a very different epoch, also of interest for velocity dependent self-scattering cross sections. And in contrast to the bounds given in Table~\ref{SIcs_limits} it has a trivial dependence on the relative amount of self-interacting dark matter $F_\mrm{SIDM}^0$.\\
	The detailed analytic calculation is given in Ref.~\cite{Stiele_2010} and is confirmed by the numerical studies of Ref.~\cite{Dutta_2010}.

\section{Dark Matter Decoupling in a Self-interaction Dominated Universe}
	Another physical process that can happen during the self-interaction dominated epoch is the decoupling of the dark matter particles. Chemical decoupling occurs when the dark matter annihilation rate $\Gamma_\mrm{A}=n_\mrm{DM}\,\langle{\sigma_\mrm{A}}v\rangle$ becomes less than the expansion rate of the universe $H\propto{\varrho^{1/2}}$.
	In a self-interaction dominated universe the expansion rate is proportional to the self-interacting dark matter particle density: $H\propto{\varrho_\mrm{SI}^{1/2}}\propto{n_\mrm{SIDM}}$.\\
	So the self-interacting dark matter annihilation cross-section is independent on the particle parameters but determined by the elastic self-interaction strength:
	\begin{equation}
		\sigma_\mrm{A}^\mrm{SIDM}\approx7.45\times10^{-7}\times\frac{100\,\mrm{MeV}}{m_\mrm{SI}/\!\sqrt{\alpha_\mrm{SI}}}\,\sigma_\mrm{weak}\;,
		\label{eq:cs_SIDM}
	\end{equation}
	with $\sigma_\mrm{weak}\approx1.24\times10^{-39}\,\mrm{cm^2}$.
	Hence, super-weak inelastic interactions are predicted by strong elastic dark matter self-interactions.
	The scaling $\sigma_\mrm{A}^\mrm{SIDM}\propto1/\left({m_\mrm{SI}\,/\!\sqrt{\alpha_\mrm{SI}}}\right)$ complies qualitatively with the statement of Ref.\ \cite{Hui_2001} ``that the elastic scattering cross section cannot be arbitrarily small given a nonvanishing inelastic cross section''.\\

\begin{figure}
		\centering
		\includegraphics[width=0.75\textwidth]{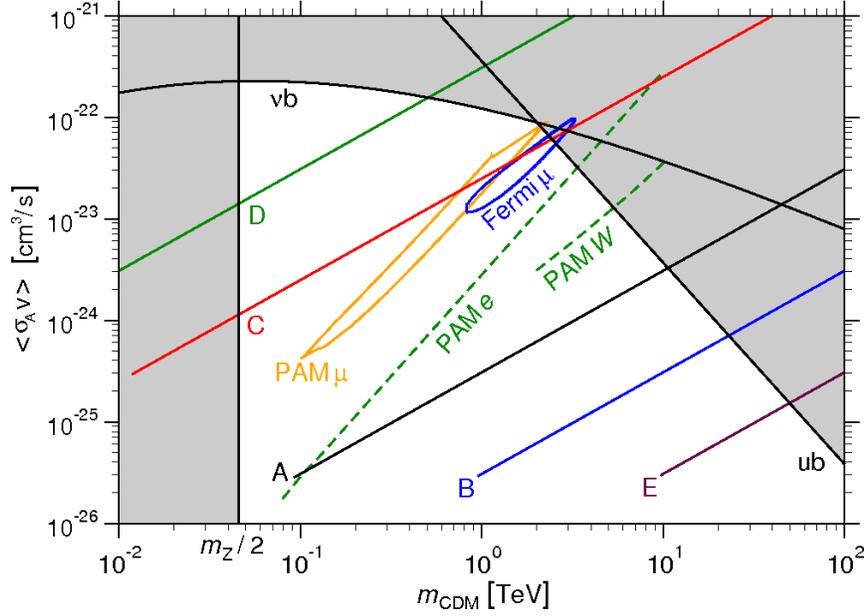}
		\caption[]{Collisionless cold dark matter annihilation cross-section in dependence of the cold dark matter particle mass
		, according to Eq.\ (\ref{eq:nat_scale}). Also shown are as upper limits the \emph{Halo Average} neutrino bound of Ref.\ \cite{Yueksel_2007} (${\nu}$b) and the unitarity bound  according to $\langle\sigma_\mrm{A}\mrm{v}\rangle\le4\pi/\left(m_\mrm{CDM}^2\,\mrm{v}\right)$ (ub, \cite{Griest_1990, Hui_2001}), as well as the $2\sigma$ contours for fits to Fermi (Fermi\,$\mu$) and PAMELA (PAM\,$\mu$) data assuming annihilation only to $\mu^+\mu^-$ of Ref.\ \cite{Bergstroem_2009} and the best-fit lines to the PAMELA data for annihilations to $e^+e^-$ (PAM\,$e$) and $W^+W^-$ (PAM\,$W$) of Ref.\ \cite{Catena_2010}. The solid lines are for the following dark matter particle parameter sets:\\

			\begin{tabular}{llll}
				\hskip17.4ex A\phantom{AA} & $m_\mrm{SI}/\!\sqrt{\alpha_\mrm{SI}}=1\,\mrm{MeV}$ & $m_\mrm{SIDM}=1\,\mrm{keV}$ & $F_\mrm{SIDM}^0=0.1$\\
				\hskip17.4ex B & $m_\mrm{SI}/\!\sqrt{\alpha_\mrm{SI}}=1\,\mrm{MeV}$ & $m_\mrm{SIDM}=10\,\mrm{keV}$ & $F_\mrm{SIDM}^0=0.1$\\
				\hskip17.4ex B & $m_\mrm{SI}/\!\sqrt{\alpha_\mrm{SI}}=10\,\mrm{MeV}$ & $m_\mrm{SIDM}=1\,\mrm{keV}$ & $F_\mrm{SIDM}^0=0.1$\\
				\hskip17.4ex C & $m_\mrm{SI}/\!\sqrt{\alpha_\mrm{SI}}=1\,\mrm{MeV}$ & $m_\mrm{SIDM}=1\,\mrm{keV}$ & $F_\mrm{SIDM}^0=0.9$\\
				\hskip17.4ex D & $m_\mrm{SI}/\!\sqrt{\alpha_\mrm{SI}}=1\,\mrm{keV}$ & $m_\mrm{SIDM}=1\,\mrm{keV}$ & $F_\mrm{SIDM}^0=0.1$\\
				\hskip17.4ex E & $m_\mrm{SI}/\!\sqrt{\alpha_\mrm{SI}}=100\,\mrm{MeV}$ & $m_\mrm{SIDM}=1\,\mrm{keV}$ & $F_\mrm{SIDM}^0=0.1$\\
				\hskip17.4ex E & $m_\mrm{SI}/\!\sqrt{\alpha_\mrm{SI}}=10\,\mrm{MeV}$ & $m_\mrm{SIDM}=10\,\mrm{keV}$ & $F_\mrm{SIDM}^0=0.1$\\
			\end{tabular} }
		\label{fig:sigmaAv_mCDM}
	\end{figure}

	An interesting possibility is the assumption of a second, collisionless cold dark matter component, usually represented by WIMPs. The natural scale of its annihilation cross-section for decoupling in a self-interaction dominated universe becomes
	\begin{eqnarray}
		\langle\sigma_\mrm{A}\mrm{v}\rangle_\mrm{CDM}&\approx&2.77\times10^{-24}\,\mrm{cm^3\,s^{-1}} \nonumber\\
		&&\times\,\frac{m_\mrm{CDM}/10\,\mrm{TeV}}{m_\mrm{SIDM}/1\,\mrm{keV}}\,\frac{10\,\mrm{MeV}}{m_\mrm{SI}/\!\sqrt{\alpha_\mrm{SI}}}\,\frac{F_\mrm{SIDM}^0}{1-F_\mrm{SIDM}^0}\;.
		\label{eq:nat_scale}
	\end{eqnarray}
	Hence, the natural scale of the cold dark matter annihilation cross-section depends on the SIDM elastic self-interaction strength and linearly on the CDM particle mass. All in all the natural scale of cold dark matter decoupling can be increased by some orders of magnitude compared to decoupling during radiation domination, in which $\langle\sigma_\mrm{A}\mrm{v}\rangle_\mrm{CDM}\sim3\times10^{-26}\,\mrm{cm^3\,s^{-1}}/\left(1-F_\mrm{SIDM}^0\right)$. This is in contrast to the \textquoteleft{WIMP miracle}' (meaning that they are not that weakly interacting). Interestingly enough, such boosted cold dark matter annihilation cross-sections are able to explain the high energy cosmic-ray electron-plus-positron spectrum measured by Fermi-LAT and the excess in the PAMELA data on the positron fraction (see e.g.\ \cite{Bergstroem_2009}).
	Fig.~\ref{fig:sigmaAv_mCDM} shows the collisionless cold dark matter annihilation cross-section for various parameters of the self-interacting dark matter and collisionless cold dark matter together with the fits to Fermi and PAMELA data. They are compatible with each other.\\
	
	{\noindent}Further calculations  and the corresponding discussions are given in Ref.~\cite{Stiele_2010}.
	
\section{Structure Formation in a Self-interaction Dominated Universe}
	Another consequence of an early self-interaction dominated epoch may concern structure formation. In the standard model dark matter structures do not grow before matter domination. But a relativistic analysis of ideal fluid cosmological perturbations reveals for self-interaction dominated self-interacting dark matter the following evolution of the density contrast $\delta$ in the subhorizon limit:
	\begin{equation}
		\delta_{\mrm{SIDM}} \propto a \cdot \left(A \cos(a^2 - 3\pi/4) + B \sin(a^2 - 3\pi/4)\right)\;,
	\end{equation}
	i.e.\ an oscillation with a linearly growing amplitude \cite{Hwang_1993}. However, any increase in the density contrast of self-interacting dark matter will be washed out either by collisional self-damping or by free streaming.\\

	A subdominant collisionless cold dark matter component allows for an increase in the density fluctuations:
	\begin{equation}
		\delta_{\mrm{CDM}} = a \cdot \left(C/{a^\mrm{in}_{k}}^2\right) + D\;,
	\end{equation}
	where $a^{\mrm{in}}_{k}$ is the scale parameter at horizon entry.
	This means that subhorizon collisionless cold dark matter density fluctuations will also grow linearly during self-interaction domination. Thus fluctuations at low masses in the matter power spectrum are enhanced. They are limited by the comoving wavenumber that is equal to the Hubble scale at self-interaction--radiation equality, corresponding to $\sim1.4 \times 10^{-3} M_\odot$ being the largest structures that can be enhanced.\\
	
	{\noindent}For the calculations and details we refer to Ref.~\cite{Stiele_2010}.

\begin{acknowledgments}
We thank Gabrijela Zaharijas, Hasan Y\"{u}ksel, Andrea Macci\`{o}, Riccardo Catena and John Beacom for fruitful discussions and for providing us with their respective data.\\
This work was supported by the German Research Foundation (DFG) within the framework of the excellence initiative through the Heidelberg Graduate School of Fundamental Physics (HGSFP) and through the Helmholtz Graduate School for Heavy-Ion Research (HGS-HIRe) and the Graduate Program for Hadron and Ion Research (GP-HIR) by the Gesellschaft f\"ur Schwerionenforschung (GSI), Darmstadt.
\end{acknowledgments}


\end{document}